\documentclass[%
 reprint,
 amsmath,amssymb,
 aps,
]{revtex4-2}

\usepackage{graphicx}
\usepackage{dcolumn}
\usepackage{bm}

\begin{document}

\preprint{APS/123-QED}

\title{ChatGPT in the Classroom: Boon or Bane for Physics Students' Academic Performance?}
\thanks{A footnote to the article title}%

\author{Manuel G. Forero }
 \altaffiliation[Also at ]{Semillero Lún, Facultad de Ingenier\'{i}a, Universidad de Ibagu\'{e}, Ibagu\'{e} 730002, Colombia}
\author{H. J. Herrera-Su\'arez}%
 \email{hernan.herrera@unibague.edu.co}
\affiliation{%
 Semillero Lún, Facultad de Ciencias Naturales y Matem\'{a}ticas, Universidad de Ibagu\'{e}, Ibagu\'{e} 730002, Colombia}%

\date{\today}

\begin{abstract}
This study investigates the influence of ChatGPT, an AI-based language model, on student performance in a physics course. We conducted an experimental analysis with two cohorts of students in a second-semester engineering physics course. The control group (Physics 2 2022B) used traditional teaching methods, while the experimental group (Physics 2 2023A) integrated ChatGPT as a learning tool. Our results indicate that the use of ChatGPT led to a significant decrease in student performance, as evidenced by lower grades and negative Hake factors compared to the control group. In addition, a survey of students revealed conflicting perceptions of the usefulness of ChatGPT in teaching physics. While most recognized its usefulness in understanding concepts and providing information, concerns were raised about its potential to reduce critical thinking and independent learning. These findings suggest that while ChatGPT can be a useful tool, it should be used with caution and as a supplement to traditional teaching methods, rather than as a stand-alone solution. The study underlines the importance of critical and reflective use of AI tools in educational settings and highlights the irreplaceable role of teachers in providing comprehensive educational support.

\end{abstract}

\maketitle


\section{\label{Introduction}First-level heading:\protect\\ }

Artificial Intelligence (AI) has experienced rapid development in this century due to its significant contributions in various fields such as task automation \cite{Brynjolfsson2014}, advancements in healthcare \cite{Topol2019, Esteva2017, Rajkomar2019, Chartrand2017}, autonomous driving \cite{shalev2017, Seff2015}, cybersecurity  \cite{Chen2020, Buczak2015}, e-commerce \cite{Bawack2022, Zhang2021, Micu2021}, weather prediction \cite{Hu2016, Ham2019}, virtual assistants, and chatbots \cite{Goldberg2022, Chen2017, Serban2016, Banchs2012}, among others. Within virtual assistants and chatbots, on September 30, 2022, OpenAI released "ChatGPT" \cite{chatgpt2022}, which is being widely used by the community, particularly by students. ChatGPT is a language model based on GPT (Generative Pre-trained Transformer) that performs conversational tasks. GPT is a language model architecture that utilizes a trained neural network to generate relevant and coherent text sequences based on a prompt.

There are different positions regarding the use of ChatGPT in education. Some believe that ChatGPT can harm education \cite{Maya2023}, while others think the opposite \cite{Adiguzel2023, Sun2023}  and others think that the two previous positions  \cite{Rahman2023}. Throughout education, the existence of other tools, such as the use of calculators, has been observed. Initially, many educators, including teachers and university professors, prohibited their use \cite{Pendleton1975}. However, over time, there has been a total acceptance of calculators within the educational community. Nowadays, educators have adapted to the changing landscape and recognize the value of calculators as a beneficial tool for students. Nonetheless, the integration of calculators into the educational system continues to spark discussions regarding their impact on student's cognitive development and problem-solving abilities. The use of calculators allows for time-saving, and ChatGPT offers similar advantages. However, we need to address the question of whether it is truly contributing to or deteriorating the quality of education. If it is the latter, it should serve as a basis for seeking new alternatives or new ways to use this tool for the benefit of students.

Given that there are very few studies available to assess how student learning and performance are affected by the use of new AI-based technology like ChatGPT \cite{Küchemann2023}, it becomes necessary to conduct this type of research in order to later establish strategies that allow us to make the most of these tools in a way that benefits student learning.

Therefore, in this research, an experiment has been conducted to see how much performance is affected by the use of ChatGPT, and if students get better or worse performance. The first population was taken as the control group and the traditional (Physics 2 2022B) method is used where the evaluations are done in a normal way and the use of ChatGPT in class is prohibited, in the second population the use of ChatGPT is allowed and they were promoted to use it (Physics 2 2023A), the result we found in this first experiment is that the performance of students was harmed notoriously by using ChatGPT, so new strategies are required to not affect the performance of students.

This paper is organized as follows: In Section 2, the methodological details are presented. In Section 3, the main results are described, and the discussion is provided in Section 4. Finally, Section 5 presents some concluding remarks.

\section{Materials and Methods}
For this research endeavor, two distinct cohorts of students participating in a second-semester engineering physics course were chosen.

Each student cohort was assigned a set of three tasks encompassing the study of theoretical concepts related to oscillations, fluids, and thermodynamics—foundational topics integral to the standard curriculum of the Faculty of Engineering's Physics II course. The age range of the students fell between 17 and 20 years.

The initial cohort, identified as the control group (Physics 2 2022B), comprised 40 students who had completed the course prior to the implementation of ChatGPT, and consequently, were unacquainted with the tool's functionalities.

The control group employed conventional pedagogical methods commonly employed at the University for physics instruction. This approach involved the dissemination of knowledge through lectures, utilizing tools such as blackboards, slides, videos, and simulation software for the analysis of physical systems, as well as laboratory sessions for practical verification of theoretical concepts. Moreover, lectures and assignments were an integral part of this pedagogical approach. Within this teaching paradigm, students followed a study methodology that included the use of the core textbook, supplemented by materials recommended in the course syllabus, and enriched by video resources accessible via platforms like YouTube.

In contrast, the second group, representing the first semester of 2023 (Physics 2 2023A), comprising 36 students, was exposed to our experimental methodology. This novel approach incorporated ChatGPT as an instructional aid within the course. This group followed a similar instructional methodology, yet they were introduced to ChatGPT 3.5 from the outset of the course. Students were taught how to utilize this tool and strongly encouraged to leverage it throughout the semester. Consequently, students were motivated to employ ChatGPT for inquiries related to course topics, crafting theoretical questions and problems, as well as seeking clarification. Periodic verifications of their interactions with ChatGPT were conducted to ensure genuine utilization of the tool.

In the evaluation process, the lecturer crafted a questionnaire aligned with the course content, featuring questions ranging from true/false definitions to open-ended inquiries that required students to demonstrate their understanding of the subject matter.

To maintain objectivity in the comparison, the same lecturer delivered lectures on identical topics, within the same time frame, and in the same sequence for both cohorts. The only variation was the additional time allotted for introducing the ChatGPT 3.5 tool.

To complement the information acquired through the experiment, a survey was conducted among 41 students enrolled in Physics courses I, II, and III, all of whom were instructed by the same teacher. These students were instructed to employ ChatGPT for the preparation of specific theoretical points that would be assessed in upcoming exams. To facilitate this, they were introduced to the utilization of ChatGPT for addressing the theoretical inquiries within their respective courses and were consistently encouraged to incorporate it into their study routines. Frequent reminders were provided during each class to reinforce the importance of utilizing ChatGPT for theory comprehension.

The primary objective of this survey was to gather insights from the students regarding the utility of ChatGPT and to gauge whether its integration had positively influenced their learning experience.

\section{Results}
The primary objective of this study was to assess the impact of ChatGPT on student performance in the academic context.

This section presents the results obtained from the research. As previously mentioned, the primary objective of this study was to experiment to examine the extent to which students' performance in a physics course (Physics 2) is influenced by the utilization of ChatGPT. To achieve this, two distinct groups were carefully selected: the first group served as the control (Physics 2 2022B), while the second group had access to ChatGPT (Physics 2 2023A). The evaluation of the Physics 2 course encompassed three examinations distributed across three grading periods.

\begin{figure}
\includegraphics[width=8 cm]{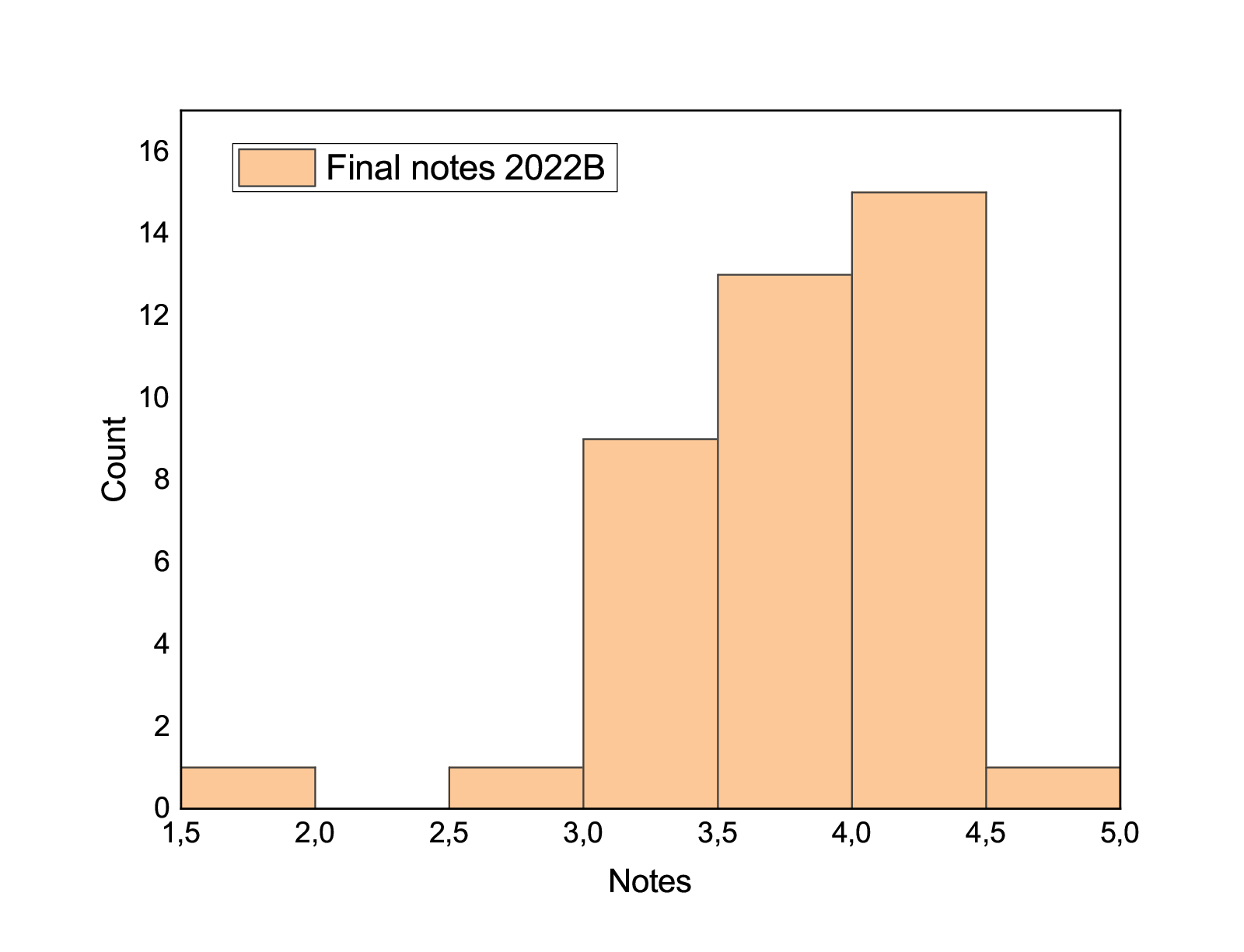}
\caption{Histogram of the distribution of grades in the control group.\label{fig:fig1}}
\end{figure}

\begin{figure}
\includegraphics[width=8 cm]{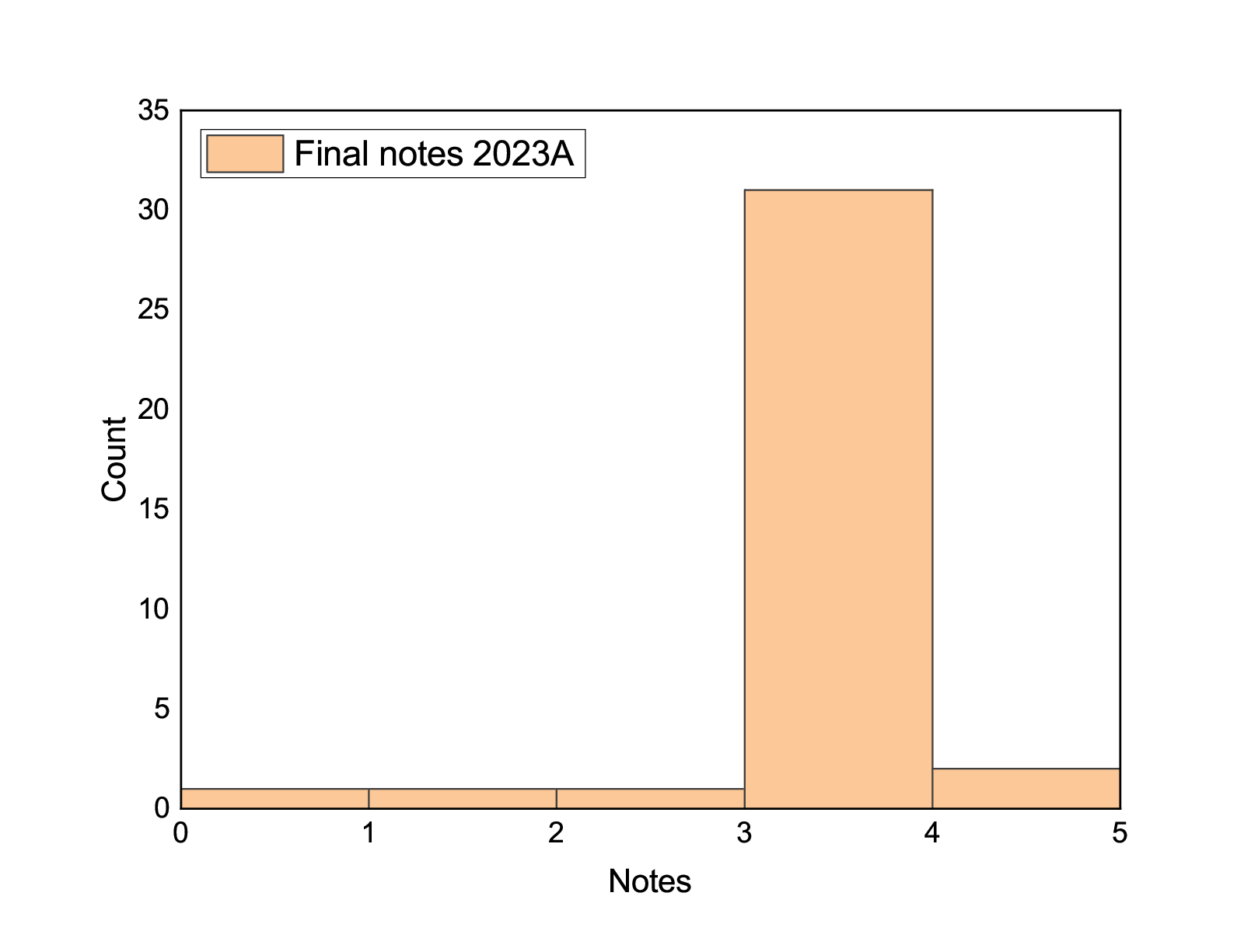}
\caption{ Histogram of Grades for the Group with Access to ChatGPT\label{fig:fig2}}
\end{figure}

In Fig.~\ref{fig:fig1}, the histogram displays the final grades of the control group students. The histogram is skewed to the right, indicating that 95\% of the students passed the subject with a grade higher than 3, while 5\% failed the subject with a grade lower than 3. Similarly, Fig.~\ref{fig:fig2} illustrates the histogram of final grades for the students in the Physics 2 2023A group, with 91.66\% of the students passing the subject with a grade higher than 3, and 8.33\%failing the subject.

\unskip
\vspace{0.5 cm}

\unskip
\vspace{0.5 cm}

Table~\ref{tab:table1} displays the mean scores per grading period for the subject, considering the two aforementioned cohorts. A comparison of the averages in columns 2 and 3 reveals that the performance of the first cohort (Physics 2 2022B) consistently outperforms that of the subsequent cohort (Physics 2 2023A) across all grading periods and in the overall aggregate.

\begin{table}[b]
\caption{\label{tab:table1}%
 Averages per period.\label{table1}
}
\begin{ruledtabular}
\begin{tabular}{lcdr}
\textrm{Physics 2  2022B}&
\multicolumn{1}{c}{\textrm{Physics 2  2023A with GPT}}\\
\colrule
Period 1		& 3.32			& 3.02\\
Period 2		& 3.94			& 3.68\\
Period 3		& 4.04			& 3.35\\
Total subject	& 3.79			& 3.35 \\
\end{tabular}
\end{ruledtabular}
\end{table}

In the first column of Table~\ref{tab:table2}, it can be observed the breakdown of assessments per grading period. Each of these assessments contributes to 20\% of the total course grade. The second column presents the average scores of the assessments for the control group (Physics 2 2022B), while the third one displays the average scores of the assessments for the group granted permission to use ChatGPT (Physics 2 2023A). Upon comparing columns 2 and 3, it becomes evident that the control group consistently outperforms the group allowed to utilize ChatGPT. 

\begin{table}[b]
\caption{\label{tab:table2}%
Evaluation averages.\label{table2}
}
\begin{ruledtabular}
\begin{tabular}{lcd}
\textrm{}&
\textrm{Physics 2  2022B}&
\multicolumn{1}{c}{\textrm{Physics 2  2023A with GPT}}\\
\colrule
Evaluation 1		& 2.88			& 2.21 \\
Evaluation 2		& 3.67			& 3.47 \\
Evaluation 3		& 3.60			& 2.78 \\
\end{tabular}
\end{ruledtabular}
\end{table}

The Hake factor is a metric used to assess conceptual learning gain in the field of education, as introduced by Hake in 1998 \cite{Hake1998}. This factor is computed using the following formula:

\begin{equation}
g=\frac{\left\langle S_{f}\right\rangle -\left\langle S_{i}\right\rangle }{%
100-\left\langle S_{i}\right\rangle }  \label{1}
\end{equation}

where $\left\langle S_{f}\right\rangle $ represents the average of the final grades, and $\left\langle S_{i}\right\rangle $ represents the average of the initial grades. Table~\ref{tab:table3} displays the assessments per grading period in its first column and the corresponding Hake factor in the second one. It is noteworthy that the Hake factor is negative for each assessment, signifying that the students from the second population, who utilized ChatGPT, have exhibited a decline in conceptual understanding after the course, as compared to the control group.

\begin{table}[b]
\caption{\label{tab:table3}%
 Hake factor for evaluation.\label{table3}
}
\begin{ruledtabular}
\begin{tabular}{lc}
\textrm{}&
\multicolumn{1}{c}{\textrm{Hake factor}}\\
\colrule
Evaluation 1		&-0.31 \\
Evaluation 2		& -0.15\\
Evaluation 3		& -0.58 \\
\end{tabular}
\end{ruledtabular}
\end{table}

The two-sample t-test was conducted, and the results are presented in  Tables~\ref{tab:table4} and ~\ref{tab:table5}. Table~\ref{tab:table4} provides information about the respective populations, including sample size, mean, standard deviation, standard error of the mean, and median.

Upon inspection of Table~\ref{tab:table4}, it is evident that the control population had a higher average grade than the population that utilized ChatGPT. Furthermore, the difference in means and medians between the two populations amounts to 0.44 and 0.38625, respectively.

Table ~\ref{tab:table5} displays the t-test results for the control group and the group that can utilize ChatGPT, in the second column for both populations. For both homogeneous and non-homogeneous variances, the t-statistic is statistically significant, indicating that the observed difference between the means of the two populations is unlikely to be due to chance. The third column presents the degrees of freedom (DF) for homogeneous and non-homogeneous variances, indicating that there is sufficient information and an adequate sample size to obtain reliable results. The fourth column shows the probability of obtaining a difference in the means of the two populations equal to or more extreme than the observed difference (Prob). For homogeneous variances, the probability is 0.00518, which is less than 0.05, providing statistical evidence that the means of the two populations are different. For non-homogeneous variances, the probability is 0.00613, still indicating that the observed difference between the means of the two populations is statistically significant.

\begin{table*}
\caption{\label{tab:table4} Descriptive statistics.}
\begin{ruledtabular}
\begin{tabular}{cccccc}
\textrm{}&
\multicolumn{1}{c}{\textrm{N}} & {\textrm{Mean}}
&{\textrm{SD}} & {\textrm{SEM}} &{\textrm{Median}}\\ \hline
Final grades 2022B & 40	& 3.79 & 0.56522  & 0.08937 & 3.87\\
Final grades 2023A & 36	& 3.35  & 0.77111  &0.12852 & 3.48375\\
\end{tabular}
\end{ruledtabular}
\end{table*}

\begin{table*}
\caption{\label{tab:table5} Descriptive statistics.}
\begin{ruledtabular}
\begin{tabular}{cccc}
\textrm{}&
\multicolumn{1}{c}{\textrm{t Statistic}} & {\textrm{DF}}
&{$ Prob > \left | t \right |$}\\ \hline
Equal variance assumed & 2.88133	& 74 & 0.00518  \\
Equal variance not assumed (Welch correction) & 2.83542	& 63.67178  & 0.00613  \\
\end{tabular}
\end{ruledtabular}
\end{table*}

As mentioned above, a survey was conducted targeting a student population of 41 people. The first question asked was: "To what extent do you consider the use of ChatGPT during the course led you to stop thinking and rely solely on the provided answers?"

\begin{figure}
\includegraphics[width=8 cm]{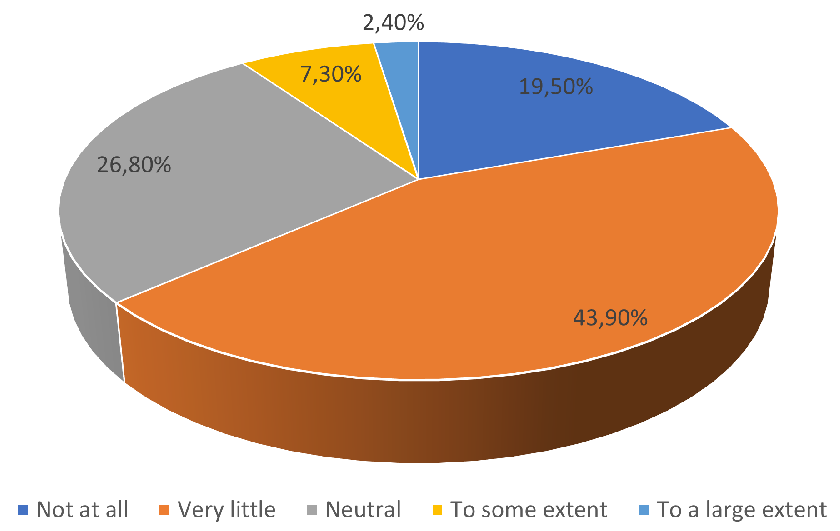}
\caption{ Distribution of responses to the question: To what extent do you consider the use of ChatGPT during the course led you to stop thinking and rely solely on the provided answers? a) Not at all b) Very little c) Neutral d) To some extent e) To a large extent .\label{fig:dc1}}
\end{figure}

Fig.~\ref{fig:dc1}  shows that $26.8\%$ of the students had a neutral stance on the question of whether ChatGPT had led them to think less critically or independently. While 9.7\% of the respondents reported that ChatGPT had affected them to a certain or large extent in this aspect, 63.4\% of students reported that ChatGPT had not made them less critical or independent thinkers.

\begin{figure}
\includegraphics[width=8 cm]{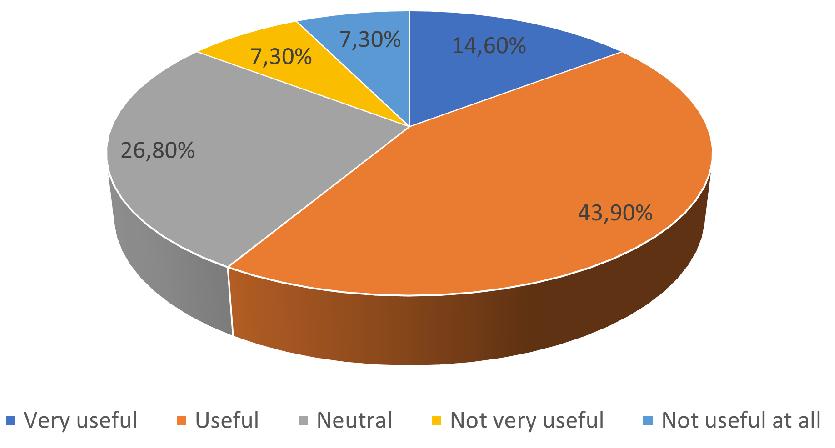}
\caption{ Distribution of responses to the question: Do you find ChatGPT's ability to answer physics-related questions useful during the course? a) Very useful b) Useful c) Neutral d) Not very useful e) Not useful at all.\label{fig:dc2}}
\end{figure}

The second question posed was: "Do you find ChatGPT's capacity to address physics-related queries beneficial during the course?" As shown in Fig.~\ref{fig:dc2}, a significant majority of students find the information provided by ChatGPT to be advantageous. A combined total of 58.5 \% of respondents categorize the tool as either "Very Useful" (14.6 \%) or "Useful" (43.9 \%), highlighting a general consensus on ChatGPT's utility for answering physics-related questions. Conversely, a combined 14.6 \% of participants rate the tool as "Not Very Useful" or "Not at All Useful," representing a smaller, yet notable, segment that does not find ChatGPT beneficial for this purpose. Additionally, 26.8 \% of respondents remain neutral on ChatGPT's usefulness in this context, highlighting an area for potential exploration and enhancement in the tool's functionality and relevance for answering physics questions.

\unskip
\vspace{0.5 cm}

\begin{figure}
\includegraphics[width=8 cm]{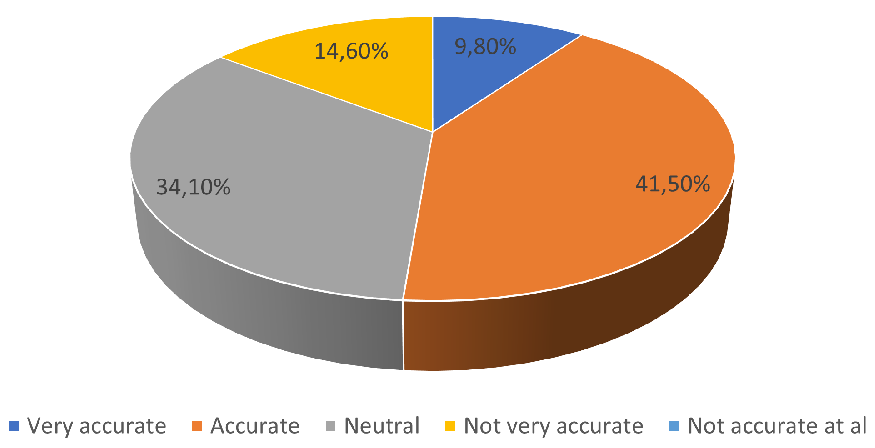}
\caption{ Distribution of responses to the question: How accurate do you find the information provided by ChatGPT regarding the studied physics concepts to be? a) Very accurate b) Accurate c) Neutral d) Not very accurate e) Not accurate at all.\label{fig:dc3}}
\end{figure}

The third survey question asked students to assess the perceived accuracy of the information provided by ChatGPT concerning the physics concepts studied. The data collected from this question is illustrated in Fig.~\ref{fig:dc3}. As shown, 9.8 \% of respondents believe that the information furnished by ChatGPT is highly accurate. Additionally, 41.5 \% of the survey participants deem the information to be accurate, while 34.1 \% hold a neutral stance on its accuracy. Conversely, 14.6 \% of respondents consider the information provided by ChatGPT to be inaccurate.

\unskip
\vspace{0.5 cm}

\begin{figure}
\includegraphics[width=8 cm]{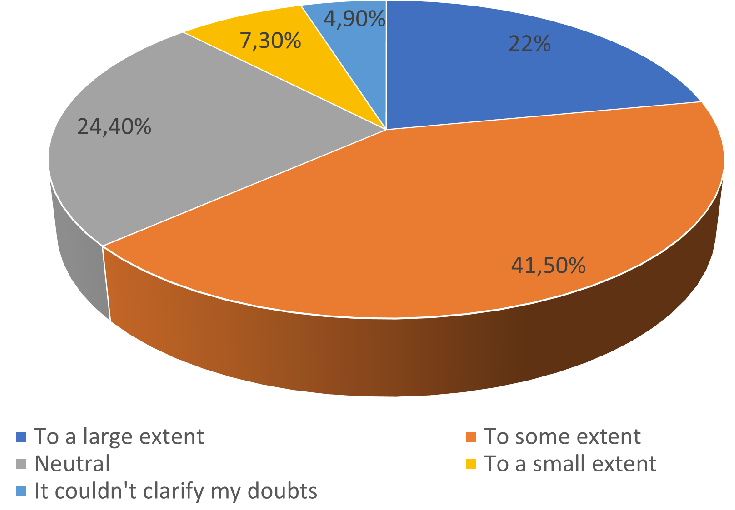}
\caption{ Distribution of responses to the question: To what extent did ChatGPT clarify your doubts and provide understandable explanations about the covered physics topics? a) To a large extent b) To some extent c) Neutral d) To a small extent e) It couldn't clarify my doubts.\label{fig:dc4}}
\end{figure}

The fourth question asked students to rate the extent to which ChatGPT could elucidate their doubts and provide clear explanations on discussed physics topics. The corresponding results are depicted in Fig.~\ref{fig:dc4}. A substantial majority (63.5 \%) of students affirmed ChatGPT's effectiveness in this area, with 22.0 \% and 41.5 \% indicating "To a Large Extent" and "To a Certain Extent," respectively. This consensus underscores a general satisfaction with ChatGPT's ability to clarify uncertainties regarding the physics topics covered. A neutral stance was maintained by 24.4 \% of participants, suggesting ambivalence or lack of engagement with the tool for this purpose. On the other hand, a combined 12.2 \% of respondents, categorized under "To a Small Extent" (7.3 \%) and "Couldn't Clarify My Doubts" (4.9 \%), expressed dissatisfaction, indicating an area where ChatGPT could potentially enhance its explanatory capabilities and doubt-resolution effectiveness.

\begin{figure}
\includegraphics[width=8 cm]{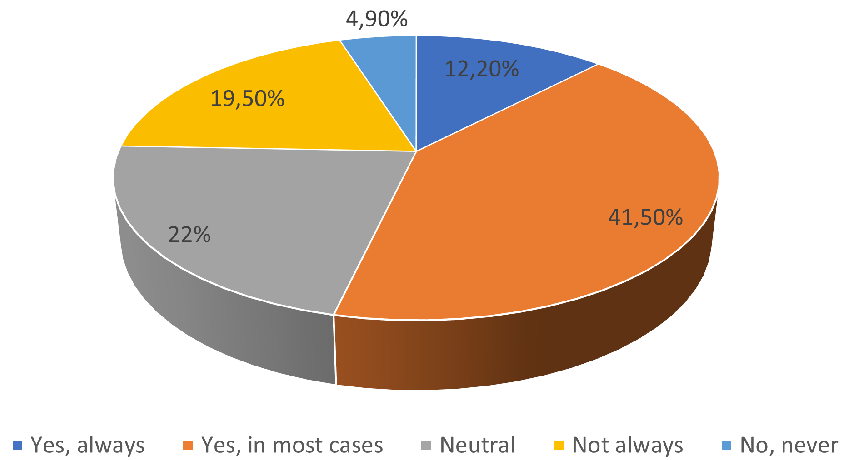}
\caption{ Distribution of responses to the question: Do you think ChatGPT was able to consistently provide relevant examples and practical applications of the studied physics concepts? a) Yes, always b) Yes, in most cases c) Neutral d) Not always e) No, never.\label{fig:dc5}}
\end{figure}

The fifth question asked students whether they think ChatGPT was able to provide relevant examples and practical applications of the studied physics concepts. Regarding this question, a minority (12.2\%) of respondents believe that ChatGPT consistently provides relevant examples and practical applications of studied physical concepts (see Fig.~\ref{fig:dc5}). Moreover, the majority of respondents (41.5 \%) indicated that they generally find ChatGPT to be effective in delivering relevant examples and practical applications. This represents the largest segment of the survey population and underscores the platform's overall efficacy. In line with this train of thought, the "Neutral" category, comprising 22.0 \% of respondents, exhibits neither a favorable nor an unfavorable stance toward ChatGPT's effectiveness in the discussed context. Conversely, the "Not Always" segment, which accounts for approximately 19.5 \% of respondents, expresses the view that ChatGPT's effectiveness in the specified context is inconsistent. This group represents nearly one-fifth of the total respondents and suggests areas where ChatGPT could improve its performance. Finally, a scant 4.9 \% of respondents fall into the "No, Never" category, firmly asserting that ChatGPT fails to provide relevant examples and practical applications in the discussed context.

\unskip
\vspace{0.5 cm}

\unskip
\vspace{0.5 cm}

\begin{figure}
\includegraphics[width=8 cm]{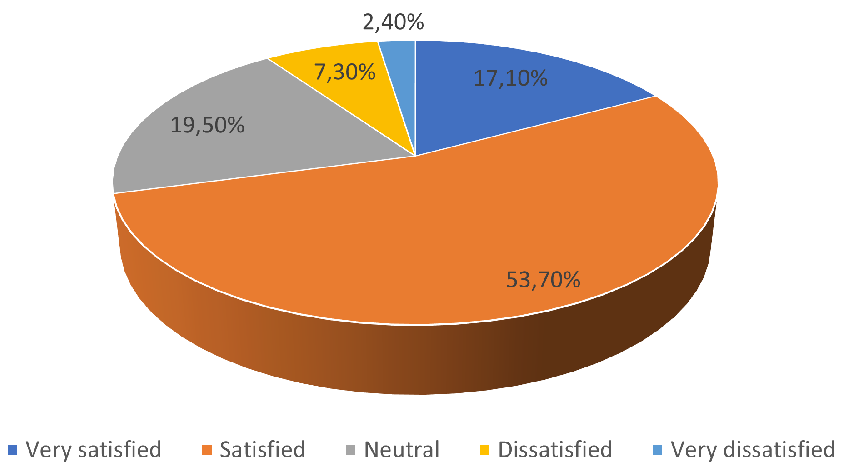}
\caption{Distribution of responses to the question: How satisfied are you with ChatGPT's ability to adapt to your level of knowledge and respond to your specific needs regarding physics? a) Very satisfied b) Satisfied c) Neutral d) Dissatisfied e) Very dissatisfied.\label{fig:dc6}}
\end{figure}

The sixth survey question asked students to rate their satisfaction with ChatGPT's ability to adapt to their level of knowledge and respond to their specific needs in the field of physics. As Fig.~\ref{fig:dc6} illustrates, a significant majority (70.8 \%) of students are either "Very Satisfied" (17.1 \%) or "Satisfied" (53.7 \%) with ChatGPT's ability to meet their specialized needs in physics. On the other side, 19.5\% of students have a neutral stance. Importantly, 9.7\% of students, categorized as "Dissatisfied" (7.3\%) or "Very Dissatisfied" (2.4 \%), are not satisfied with ChatGPT's ability to adapt to different knowledge levels and cater to specific educational needs in physics.

\unskip
\vspace{0.5 cm}

\begin{figure}
\includegraphics[width=8 cm]{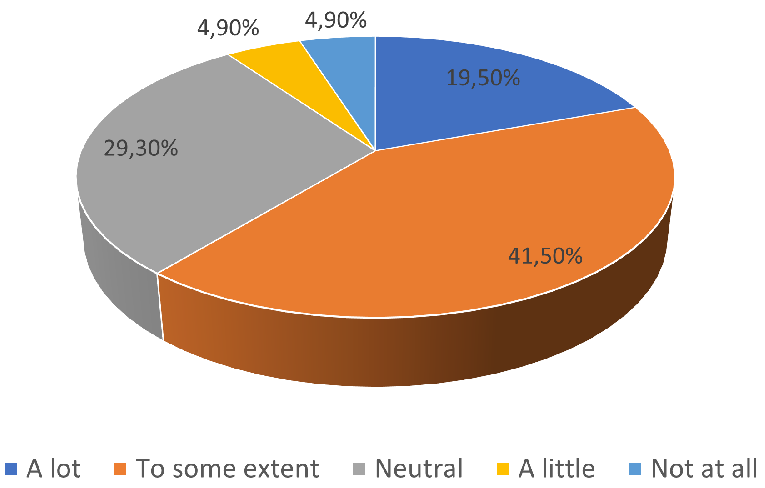}
\caption{ Distribution of responses to the question: "Do you think that interacting with ChatGPT in the course helped you deepen your understanding of the physics concepts? a) A lot b) To some extent c) Neutral d) A little e) Not at all.\label{fig:dc7}}
\end{figure}

The seventh question asked students whether they think interacting with ChatGPT in the course helped them deepen their understanding of the physics concepts. By combining the responses from the "A Lot" (19.5 \%) and "To Some Extent" (41.5 \%) categories, it is observed that a significant 61 \% of students believe that interacting with ChatGPT has been pivotal in enhancing their grasp of physics concepts, as illustrated in Fig.~\ref{fig:dc7}. Additionally, 29.3 \% of students hold a neutral opinion on this issue. Conversely, a minor segment of 9.8 \%, consisting of the "Little" (4.9 \%) and "Nothing" (4.9 \%) categories, expresses that the interaction with ChatGPT had a negligible effect on deepening their understanding of the physics concepts.

\unskip
\vspace{0.5 cm}

\begin{figure}
\includegraphics[width=8 cm]{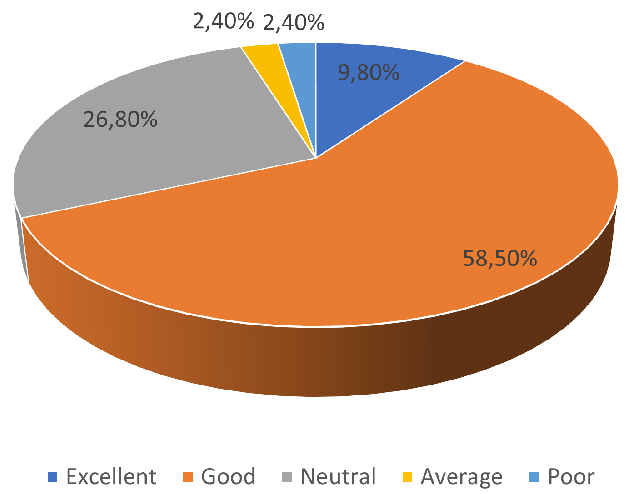}
\caption{Distribution of responses to the question: "How do you evaluate the availability and timely response of ChatGPT in addressing your queries or concerns related to physics? a) Excellent b) Good c) Neutral d) Average e) Poor. \label{fig:dc8}}
\end{figure}

The eighth question was, "How do you evaluate the availability and timely response of ChatGPT in addressing your queries or concerns related to physics?" According to Fig.~\ref{fig:dc8}, 68.3 \% of respondents hold a favorable view of ChatGPT's prompt responses. This figure is obtained by adding the percentages of respondents who rated ChatGPT's responses as "excellent" (9.8 \%) and "good" (58.5 \%). Another 26.8 \% of respondents were neutral, and the remaining 4.8 \% of respondents rated ChatGPT's responses as "average" or "poor".

\unskip
\vspace{0.5 cm}

\begin{figure}
\includegraphics[width=8 cm]{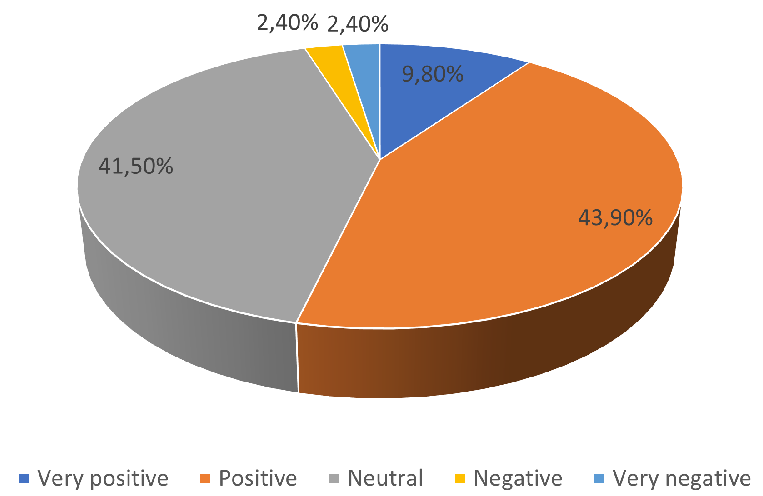}
\caption{Distribution of responses to the question: "What impact did using ChatGPT have on your motivation and active participation during the physics course? a) Very positive b) Positive c) Neutral d) Negative e) Very negative.\label{fig:dc9}}
\end{figure}

The ninth question was, "What impact did using ChatGPT have on your motivation and active participation during the physics course?" According to the results, 9.8 \% of students found that ChatGPT had a very positive impact on their motivation and active participation, 43.9 \% found that it had a positive impact, and 41.5 \% found that it had a neutral impact (see Fig.~\ref{fig:dc9}). Only a small minority of students (2.4 \%) found that ChatGPT hurt their motivation and active participation.

\unskip
\vspace{0.5 cm}

\begin{figure}
\includegraphics[width=8 cm]{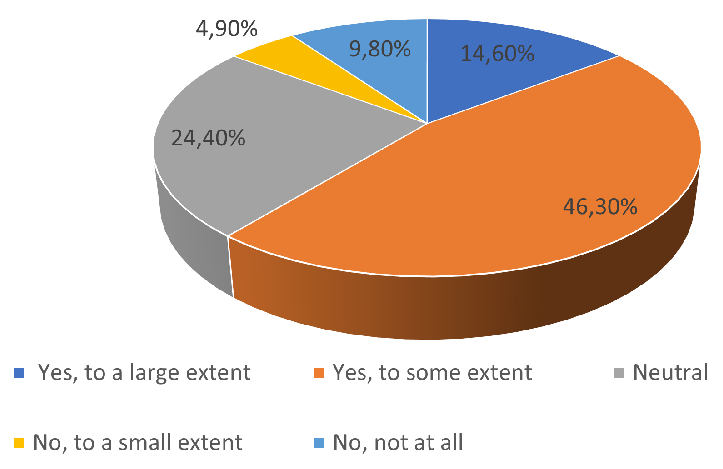}
\caption{ Distribution of responses to the question:¿Do you believe that interacting with ChatGPT in the course allowed you to explore new approaches or perspectives in understanding physics? a) Yes, to a large extent b) Yes, to some extent c) Neutral d) No, to a small extent e) No, not at all.\label{fig:dc10}}
\end{figure}

The tenth question was ¿Do you believe that interacting with ChatGPT in the course allowed you to explore new approaches or perspectives in understanding physics? Combining the categories 'Yes, to a Large Extent' (14.6 \%) and 'Yes, to Some Extent' (46.3 \%) accounts for 60.9 \% of students. This majority indicates that their interaction with ChatGPT enabled them to explore new approaches or perspectives in their understanding of Physics, as illustrated in Fig.~\ref{fig:dc10}. Conversely, approximately one-fourth of respondents fall into the 'Neutral' category. Additionally, the sum of the categories 'No, to a Small Extent' (4.9 \%) and 'No, Not at All' (9.8 \%) represents 14.7\% of respondents, suggesting that their interaction with ChatGPT did not yield new approaches or perspectives.

\unskip
\vspace{0.5 cm}

\begin{figure}
\includegraphics[width=8 cm]{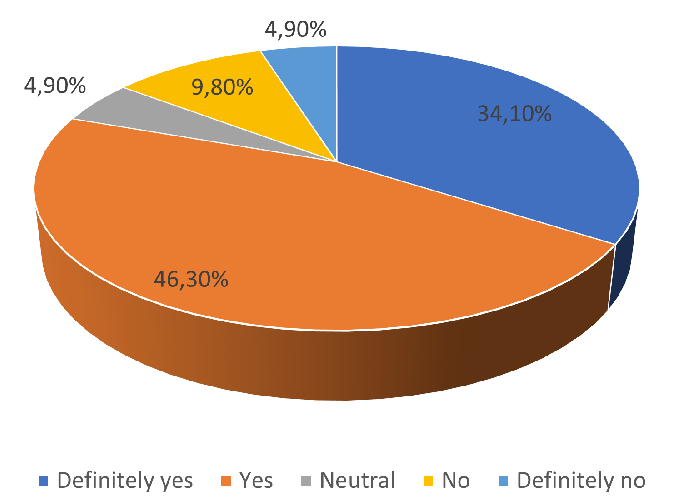}
\caption{Distribution of responses to the question: ¿Overall, would you recommend the use of ChatGPT as a support tool in physics courses to other students? a) Definitely yes b) Yes c) Neutral d) No e) Definitely no.\label{fig:dc11}}
\end{figure}

The eleventh question was: "Overall, would you recommend the use of ChatGPT as a support tool in physics courses to other students?. Combining the categories "Definitely yes" (34.1 \%) and "Yes" (46.3 \%), 80.4 \% of participants would recommend the use of ChatGPT as a support tool in physics courses. Meanwhile, 4.9 \% remained neutral (Fig.~\ref{fig:dc11}). The combined categories of "No" (9.8\%) and "Definitely no" (4.9 \%) represent 14.7 \%, a smaller percentage that would not recommend the use of ChatGPT.

\unskip
\vspace{0.5 cm}

\section{Discussion}

The results of this study suggest that the use of ChatGPT, a large language model, has a negative impact on student academic performance in a physics course. Students who used ChatGPT earned lower grades on all exams, and the measure of conceptual gain, the Hake factor, was negative for all exams. Additionally, a two-sample t-test revealed a statistically significant difference in mean grades between the two groups.

These results are consistent with the findings of another study that investigated the impact of ChatGPT on the development of high school physics assignments in grade 10, observing that students who used this broad linguistic model to learn physics had less deep learning than those who used the textbook usually employed for the course.

There are several possible explanations for the results of this study. One possibility is that ChatGPT can provide students with easy or incorrect answers to their questions, which can lead to a superficial understanding of the material. Another possibility is that ChatGPT can make students less likely to engage actively in learning, which can lead to a decrease in knowledge retention.

Regardless of the cause, the results of this study suggest that educators should be cautious when considering the use of ChatGPT or similar tools in the classroom. It is important to carefully evaluate the potential impact of these tools on student learning before implementing them.

In addition to the specific findings of this study, there are some general considerations that educators should keep in mind when evaluating the use of ChatGPT or other AI-powered learning tools. First, it is important to keep in mind that these tools are not replacements for teachers. Teachers are still essential to provide instruction, feedback, and support to students. Second, it is important to use these tools in a reflective and critical way. Educators should ensure that the tools are used in a way that promotes meaningful learning and the development of student skills.

The findings of the survey provide valuable insights into student perceptions of ChatGPT as a learning tool in a physics course. While the majority of students found ChatGPT to be a helpful resource for answering physics-related questions, clarifying doubts, and providing relevant examples, some expressed concerns about its ability to promote critical thinking and independent learning.

The survey results suggest that ChatGPT can be an effective tool for students who need help understanding physics concepts. Over 60 \% of students reported that interacting with ChatGPT helped them deepen their understanding of the material. Additionally, 68 \% of students were satisfied with ChatGPT's ability to adapt to their level of knowledge and respond to their specific needs.

However, the survey also highlights some potential drawbacks of using ChatGPT. Nearly 30 \% of students reported that using ChatGPT made them less likely to think critically or independently. Additionally, over 10\% of students were not satisfied with ChatGPT's accuracy or its ability to provide clear explanations.

Overall, the survey findings suggest that ChatGPT can be a valuable tool for students in a physics course, but it is important to use it cautiously. Students should be aware of the potential drawbacks of using ChatGPT and should use it as a supplement to, not a replacement for, traditional learning methods.

\section{Conclusions}
In conclusion, the findings of this study provide valuable insights into the potential impact of ChatGPT, a large language model, on student learning in a physics course. The study found that ChatGPT can be a useful tool for answering questions, clarifying doubts, and providing relevant examples. However, the study also found that ChatGPT can have some potential drawbacks, such as making students less likely to think critically or independently.

The study found that ChatGPT can be a valuable tool for students who need help understanding physics concepts. Over 60 \% of students reported that interacting with ChatGPT helped them deepen their understanding of the material. Additionally, 68 \% of students were satisfied with ChatGPT's ability to adapt to their level of knowledge and respond to their specific needs.

ChatGPT can also be a helpful tool for students who want to learn at their own pace. ChatGPT is permanently available, so students can access it whenever they need help. Additionally, ChatGPT can be used to personalize the learning experience by providing tailored feedback and explanations.

The study also found that ChatGPT can have some potential drawbacks. Nearly 30 \% of students reported that using ChatGPT made them less likely to think critically or independently. Additionally, over 10 \% of students were not satisfied with ChatGPT's accuracy or its ability to provide clear explanations.

It is important to be aware of these potential drawbacks when using ChatGPT. Students should use ChatGPT as a supplement to, not a replacement for, traditional learning methods. Students should also be critical of the information provided by ChatGPT and should check the accuracy of ChatGPT's answers with other sources. The study suggests that the use of ChatGPT has a negative impact on student academic performance in a physics course. Students who used ChatGPT earned lower grades on all exams, and the measure of conceptual gain, the Hake factor, was negative for all exams. These results are consistent with the findings of another study that investigated the impact of ChatGPT on the development of high school physics assignments in grade 10, observing that students who used this broad linguistic model to learn physics had less deep learning than those who used the textbook usually employed for the course.

There are several possible explanations for the results of this study. One possibility is that ChatGPT can provide students with easy or incorrect answers to their questions, which can lead to a superficial understanding of the material. Another possibility is that ChatGPT can make students less likely to engage actively in learning, which can lead to a decrease in knowledge retention.

Regardless of the cause, the results of this study suggest that educators should be cautious when considering the use of ChatGPT or similar tools in the classroom. It is important to carefully evaluate the potential impact of these tools on student learning before implementing them. Educators should ensure that the tools are used thoughtfully and critically and that they are not seen as substitutes for lecturers. Faculty remain essential in providing instruction, feedback, and support to students.
Educators should ensure that the tools are used in a reflective and critical way and that they are not seen as replacements for teachers. Teachers are still essential to provide instruction, feedback, and support to students.

Further research is needed to understand the long-term impact of using ChatGPT and other AI-powered learning tools in education. Specifically, future research could focus on areas such as the impact of ChatGPT and similar models like Bard on student performance and their impact on the development of critical thinking and independent learning skills.

\nocite{*}

\bibliography{apssamp}

\end{document}